# Finding the best density functionals for water: benchmarking the forces

Shifan Cui*

A DFT benchmark on water including more than 50 functionals from GGA to double-hybrid levels is reported. The main metric is the accuracy of forces, allowing better structural coverage, higher statistical confidence, and fewer error sources compared to conventional benchmarks. The input structures include water clusters of 4~128 molecules, with highly varied yet realistic configurations of water, ice, and their mixtures. The B97M-V, ωB97M-V, and revDOD-PBEP86-D4 functionals are found to be best on the mGGA, hybrid, and double-hybrid levels, respectively. No satisfying GGA functionals are found, but it can be obtained by combining BLYP-D4 with a simple correction to O-H bonds, achieving accuracy comparable to B97M-V.

* shifan.c@outlook.com

# 1 Introduction

Water is one of the most important substances in chemistry. On the quantum mechanics level, it is frequently modelled by density functional theory (DFT), due to the robust description on a wide range of properties at a relatively low cost. The accuracy of DFT depends on the approximate functional used and the quantity of interest, and a wise choice of functional is important for getting good results. For example, the extremely popular PBE[1] functional, despite being successful in many other fields, severely overestimates the melting point of ice[2].

Benchmarking is the canonical way to find good functionals for a particular problem. In such a process, calculations are performed with many candidate DFT functionals, and the outcomes are compared with accurate reference results.

DFT benchmarks on water have been performed on various properties, including energies[3-10], vibration frequencies[11], HOMO-LUMO gap[12], pressure of phase transition[13], density[14], melting point[2, 14-16], and spatial correlation functions[17]. Roughly speaking, these benchmarks can be categorized into two approaches:

(1) Calculating single-point, scalar quantities (e.g., energy or frequency) on a set of structures [3-13], and comparing with accurate reference methods;
(2) Carrying out molecular dynamics (MD) to calculate statistical properties (e.g. density or correlation functions)[2, 14-17], and comparing with experiments.

The first approach has the advantage that highly accurate reference results are readily available. However, the results are usually only a few scalars, which can be both positively and negatively affected by error sources. As a result, different errors may easily cancel each other, leading to fictitious good accuracy. The second approach, by introducing metrics like correlation functions, can potentially alleviate the problem. However, as accurate theoretical results are usually not available for these statistical quantities, the MD-based benchmarks have to compare with experiments instead. Furthermore, as MD is much more expensive than single-point calculations, various compensations of accuracy have to be made. Such a procedure leads to many other errors being involved, including finite size effects[18], nuclear quantum effects[19], experimental errors, and errors from force field approximations if they are used. As a result, the benchmark may not genuinely reflect the error of DFT functionals themselves.

Thanks to the increasing computing capabilities, ab-initio MD has become a popular method in materials modeling. In MD simulations, forces are the primary quantities involved. However, the error of forces themselves is not commonly evaluated in DFT benchmarks. Unlike MD statistics, forces can be calculated with both high-level reference methods and DFT methods at relatively low costs. Furthermore, the force is a 9N-dimensional quantity (where N is the number of water molecules), so it is less prone to error cancellations than in traditional scalar-based benchmarks. In other words, a much higher statistical confidence can be achieved with the same number of samples. In a water cluster, different atoms stay at different points on the potential energy surface (PES) and feel different chemical environments. With all forces calculated, the whole PES is naturally sampled, selecting out only functionals performing well in highly varied scenarios.

The objective of this paper is to benchmark DFT functionals for water, with a main focus on forces. To provide guidance for modern simulations, some relatively recent or less recognized, but still commonly available methods, are included. The remaining parts of the paper will be organized as follows. First, Section 2 describes how this benchmark is constructed and performed. Next, Section 3 shows the benchmark results. Section 4 discusses some questions related to the benchmark. Finally, Section 5 gives recommendations for choosing DFT functionals.

## 2 Methodology

### 2.1 The benchmark structure set

The first step of benchmarking is to construct the atomic structures to be evaluated. For water, this usually appears as clusters of water molecules, or crystals of ice[20]. In this work, water clusters will be used due to the high flexibility of configurations, and easier evaluation of forces. Roughly speaking, these clusters may be built in two ways:

(1) carefully made structures aiming to cover typical or important arrangements of water molecules[3, 21];

(2) structures randomly extracted from MD trajectories[7].

The second approach is closer to reality, but the sampled structures are limited to the phase simulated by MD. The first approach may allow more diverse configurations, but they could be far away from what appears in real condensed systems. In this work, a combined approach is used: MD scenes covering varied scenarios in real simulations are created, and clusters are extracted from the trajectories of these scenes.

Liquid water and ice $I_h$ are the most common phases of water in nature, and it is a good idea to have the structures covering both phases. In this work, this is achieved by sampling configurations of the premelting ice. Premelting of ice[22] happens at temperatures slightly lower than the melting point, where the surface region largely melts while the bulk still holds solid (Fig. 1). This special state of ice not only covers both solid and liquid phases, but also their surfaces and the interfaces between them.

The ratio between solid and liquid can be adjusted by tuning the temperature, creating even more diverse structures. Furthermore, this spectrum of structural order is complemented by adding pure liquid and pure ice in samples. In this way, a wide range of realistic water configurations is included, ensuring both good PES coverage and practical relevance.

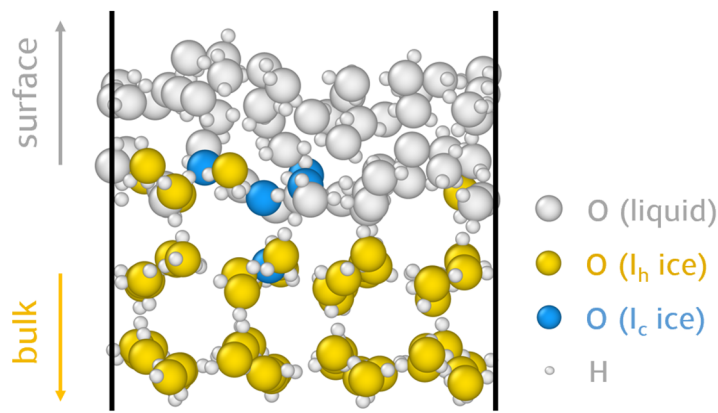

Figure 1. An example of the premelting ice, on the (0001) facet. The oxygen molecules are colored by their local structure types. The structure types are determined by an extension of the common neighbor analysis (CNA)[23]. There are more layers of molecules on the bulk side, which are not shown here. The black vertical lines are the boundaries of the simulation cell.

Another consideration is the size of clusters to be benchmarked. Smaller clusters enable more accurate reference data, but are less related to more practically relevant bulk phases. In this work, a three-level hierarchy is constructed from small to large:
(1) level 1 (Fig. 2a), containing 8 clusters, and 4 water molecules in each one;
(2) level 2 (Fig. 2b), containing 10 clusters, and 16 water molecules in each one;
(3) level 3 (Fig. 2c), containing 20 clusters, and 128 water molecules in each one.

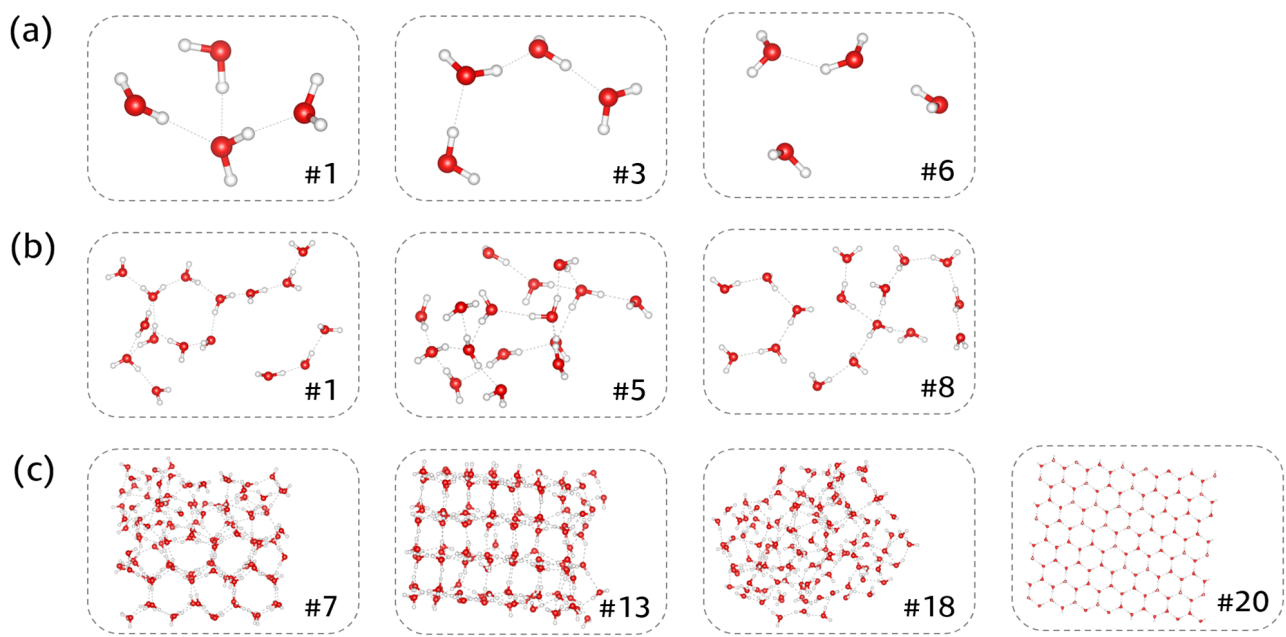

Figure 2. Examples of water clusters in level 1 (a), 2 (b), and 3 (c). The # number in each subfigure is its ID in the supplementary files. Oxygen atoms are colored red, and hydrogen atoms are colored white. Dashed lines are hydrogen bonds, which are drawn only for visualization.

Fig. 2 shows some representative clusters in each category. Most of the clusters are extracted from MD trajectories of the premelting ice, bulk ice, or bulk water. The only two exceptions are 1D and 2D configurations in level 3, for testing the asymptotic behavior at very large scales. The details of MD simulations are provided in the supporting information (SI) Section 1[24-31], and the exact coordinates of all configurations are provided in supplementary files. Only clusters of whole water molecules are considered in this work; protonated or deprotonated water is not included.

## 2.2 Computational setup

Benchmarking forces, instead of energies, introduces extra difficulties in calculating the reference data. The coupled cluster expansion up to doubles and perturbative triples (CCSD(T))[32] is the most common gold standard for ground-state energies. However, its $O(N^7)$ scaling prohibits using large basis sets that are close enough to the complete basis set (CBS) limit. Typically, this issue is circumvented by a basis set extrapolation, usually done by calculating at two adjacent basis set levels[33-35]. However, as shown in SI Section 2, this scheme cannot be directly extended to forces, and would lead to large errors otherwise.

The F12 explicit correlation method[36] is an approach allowing near-CBS results without extrapolation. By explicitly including two-body functions in the trial wavefunction, the same level of accuracy can be achieved with much fewer basis functions. As shown in SI Section 3, the results at CCSD(T)-F12/cc-pCVTZ-F12[37-39] level are very close to the regular CCSD(T) at jul-cc-pwCVQZ[40-42] or 5Z levels. Even though CCSD(T)-F12 has no analytical gradients, the PES is perfectly smooth for numerical differentiation and the result is very close to canonical CCSD(T) (Fig. S1). In this work, the CCSD(T)-F12/cc-pCVTZ-F12 level is used for calculating reference data at level 1 (4 molecules each).

For level 2 with 16 molecules each, the CCSD(T)-F12 approach also becomes computationally unfeasible. A possible solution is to run CCSD(T) with local correlation methods, such as clusters in molecules (CIM)[43], generalized energy-based fragmentation (GEBF)[44], or domain-based local pair natural orbitals (DLPNO)[45, 46]. These methods have much lower scaling with system size compared to the canonical CCSD(T), allowing for simulating larger water clusters. However, the high cost from large basis sets is still a problem. Even though DLPNO can be used together with the F12 method[37], this combination introduces extra noise in numerical differentiation, significantly affecting the accuracy (SI Section 3, Fig. S1-S2).

To solve this problem, it is useful to consider filling the basis set gap with the MP2 correlation energy difference[47]:

$$E_{corr,CCSD(T),L} = E_{corr,CCSD(T),S} + (E_{corr,MP2,L} - E_{corr,MP2,S}) \qquad (1)$$

where L and S refer to large and small basis sets, and $E_{corr}$ refers to the correlation energy of the given method. In principle, it is possible to extrapolate forces with this formula as well, by taking derivatives to atom coordinates on both sides, and multiplying by -1:

$$F_{corr,CCSD(T),L} = F_{corr,CCSD(T),S} + (F_{corr,MP2,L} - F_{corr,MP2,S}) \qquad (2)$$

where $F_{corr}$ is the "correlation correction" of forces, i.e., the CCSD(T) or MP2 forces minus the HF forces. In this work, the large basis is taken to be jul-cc-pwCVQZ while the small basis is cc-pVTZ. As shown in SI Section 3, such a scheme leads to satisfactory accuracy for water. This approach is used for level 2 clusters in this work.

For level 3 with 128 molecules each, none of the methods above are feasible anymore. Therefore, the best-performing double-hybrid density functional on level 1 and 2 (which is revDOD-PBEP86-D4[48], as it will be shown) is used as the reference method, with a more practical def2-TZVPP basis set[49] and the frozen-core approximation applied. This combination is used to benchmark rung 4 and lower functionals.

Table 1 summarizes the computational levels of all calculations in this work, with more details available in SI Section 1[50-53]. The cc-pwCVnZ basis sets[42] are used for double hybrids to describe core-valence correlations in the MP2 part, unless the frozen core approximation is used. In other cases, the def2 series[49] basis sets are used for DFT.

Table 2 lists all functionals presented in this work, categorized by the DFT rung. The list is compiled based on popularity[54], previous benchmark results on water, availability in LibXC[55], and the year of publication. The idea is to include both popular functionals and some less recognized ones, with good coverage of different rungs on the Jacobi ladder. Furthermore, the B97-3c[56] and r2SCAN-3c[57] composite methods are also included, with both their built-in small basis sets and standard large basis sets tested. The DFT-D4[58]/D3[59, 60] corrections are applied to most functionals where the parameters are available; the effect of this correction will be discussed in Section 4.1. The three-body $r^{-9}$ term is included for D4 but not for D3. For entries with the -V suffix, the non-local VV10 dispersion correction[61] is used instead of DFT-D. All quantum mechanics calculations are performed with ORCA 6.1[62], and visualizations are performed with VESTA[63] and OVITO[64].

|  | Level 1 (4H$_2$O) | Level 2 (16H$_2$O) | Level 3 (128H$_2$O) |
|---|---|---|---|
| Reference | CCSD(T)-F12 /cc-pCVTZ-F12 | DLPNO-CCSD(T1) /cc-pVTZ + DLPNO-MP2 /jul-cc-pwCVQZ | revDOD-PBEP86-D4 /def2-TZVPP with frozen core |
| Double-hybrids MP2 | aug-cc-pwCV5Z | jul-cc-pwCVQZ | - |
| Other functionals | def2-QZVPPD | def2-QZVPPD | def2-TZVPP |

*Table 1. The methods and basis sets used in this work.*

| Rung 2 (GGA) | Rung 3 (mGGA) | Rung 4 (hybrid) | Rung 5 (2-hybrid) |
|---|---|---|---|
| B97-D4[65] | B97M-V[77] | B3LYP-D4[90] | MP2 |
| BLYP-D4[66] | M06-L-D4[78] | B3PW91-D3[91] | B2PLYP-D4[109] |
| BP86-D4[67, 68] | MGGAC[79] | BHandHLYP-D4[92] | mPW2PLYP-D4[110] |
| GAM[69] | MN15L[80] | CAM-B3LYP-D4[93] | Pr2SCAN50-D4[108] |
| N12[70] | MS2[81] | HSE06-D3[94, 95] | Pr2SCAN69-D4[108] |
| OLYP-D4[71] | r2SCAN-D4[82] | LC-ωPBE-D3[96] | PWPB95-D4[111] |
| OPBE-D4[1, 71] | revM06L[83] | M05-2X[97] | revDOD-PBEP86-D4[48] |
| PBE-D4[1] | revTM[84] | M06-2X-D3-0[98] | revDSD-PBEP86-D4[48] |
| PW91-D4[72] | revTPSS-D4[85] | M06-D4[98] | ωB97X-2-D3[112] |
| revPBE-D4[73] | rSCAN-D4[86] | PBE0-D4[99] | |
| rPBE-D4[74] | SCAN-D4[87] | PW6B95-D4[100] | |
| SG4[75] | SCAN-V | r2SCAN0-D4[101] | |
| SOGGA11[76] | TM[88] | r2SCANh-D4[101] | |
|  | TPSS-D4[89] | revPBE0-D4[102] | |
|  |  | SOGGA11-X[103] | |
|  |  | TPSS0-D4[104] | |
|  |  | TPSSh-D4[89] | |
|  |  | ωB97-D4[105] | |
|  |  | ωB97M-D4[106] | |
|  |  | ωB97M-V[107] | |
| B97-3c[56] (large) | r2SCAN3c[57] (large) | ωR2SCAN-D4[108] | |
| B97-3c (ori) | r2SCAN3c (ori) |  | |

*Table 2. The DFT functionals presented in this work. 3c composite methods are tested with both the large basis sets in Table 1, and their original basis sets (def2-mTZVP or def2-mTZVPP). D3 refers to the Becke-Johnson (BJ) damping variation[60] unless explicitly stated as D3-0. MP2 is included in rung 5 for comparison.*

## 2.3 Definition of metrics

To measure the accuracy of forces, an intuitive metric is the root mean square error (RMSE) on individual atoms:

$$RMSE(F) = \sqrt{\frac{\sum_{1\leq n \leq N}\sum_{i=1,2,3}\sum_{d=1,2,3}\left|F_{n,i,d}^{ref} - F_{n,i,d}^{DFT}\right|^2}{9N}} \quad (3)$$

where $N$ is the number of water molecules, $i = 1/2/3$ represents the three atoms in a molecule, and $d = 1/2/3$ represents $x/y/z$ directions. $F_{n,i,d}$ refers to the force felt by atom $i$ in molecule $n$, in the direction $d$. The denominator $9N$ is the number of scalar force components. In this benchmark, the RMSE on every cluster is independently evaluated, and the arithmetic mean is taken between clusters at the same level.

The definition (3) itself has no problem, but it is not aware of the distinction between bonded and non-bonded interactions, which determine different aspects of physics. Therefore, two metrics of intermolecular forces are additionally defined in this work:

(1) the "rigid molecule force", which is the total force felt by a whole molecule:

$$F_{n,mol,d} = \sum_{i=1,2,3} F_{n,i,d} \quad (4)$$

Here $F_{n,mol,d}$ is the total force on molecule $n$ in direction $d$. In rigid body MD simulations (e.g., SPC/E[113] or TIP4P[114, 115]), this would be the only force of interest. Then the error of this force is defined as usual:

$$RMSE(F_{mol}) = \sqrt{\frac{\sum_{1\leq n\leq N}\sum_{d=1,2,3}\left|F^{ref}_{n,mol,d} - F^{DFT}_{n,mol,d}\right|^2}{3N}} \quad (5)$$

(2) the "force difference due to interaction", defined as following:

$$F_{n,i,d,int} = F_{n,i,d} - F_{n,i,d,iso} \quad (6)$$

where $F_{n,i,d,iso}$ is the force felt by the same atom in the same molecule, but in an isolated state (i.e., all other molecules are removed). Compared to the rigid molecule force, this metric also includes the effects of intermolecular inductions on single-atom forces. The RMSE is defined as usual:

$$RMSE(F_{int}) = \sqrt{\frac{\sum_{1\leq n\leq N}\sum_{i=1,2,3}\sum_{d=1,2,3}\left|F^{ref}_{n,i,d,int} - F^{DFT}_{n,i,d,int}\right|^2}{9N}} \quad (7)$$

Calculating it needs $F_{n,i,d,iso}$, which requires N gradient calculations, each on one water molecule. Due to the increased complexity and computational cost, this metric is not evaluated on level 3 clusters where N=128.

Besides the force metrics, the regular energy difference metric is also included in this benchmark. The error of energy is measured by the mean absolute error (MAE) per molecule, defined as

$$MAE(E) = \frac{\sum_{1\leq c\leq N_c}\left|\left|E^{ref}_c - \overline{E^{ref}_c}\right| - \left|E^{DFT}_c - \overline{E^{DFT}_c}\right|\right|}{N_c N} \quad (8)$$

where $N_c$ is the number of clusters in a level, $E_c$ is the energy of cluster c, and $\overline{E_c}$ is the average cluster energy in the level. Some previous works[6, 7, 10] used the energy of the first structure instead of the average, giving the first structure too much weight in the result compared to others. With the definition (8), each structure participates in the metric equally.

## 3 Results

The four metrics in this work ($RMSE(F), MAE(E), RMSE(F_{int}), RMSE(F_{mol})$) vary in both units and magnitude, which is inconvenient for demonstration. Therefore, they are further normalized by dividing the average error of all functionals in that metric. The normalization factors are given in Table 3. The normalized errors of all functionals are presented in Fig. 3~6, categorized by the DFT rung. The raw error numbers of all functionals are available in the supplementary files.

| Level | $RMSE(F)$ | $MAE(E)$ | $RMSE(F_{int})$ | $RMSE(F_{mol})$ |
|---|---|---|---|---|
| 1 | 0.140 | 3.784 | 0.036 | 0.021 |
| 2 | 0.144 | 3.083 | 0.051 | 0.029 |
| 3 | 0.149 | 3.038 | - | 0.028 |

*Table 3. Normalization factors of all metrics. The unit of force is eV/Å, and the unit of energy is meV.*

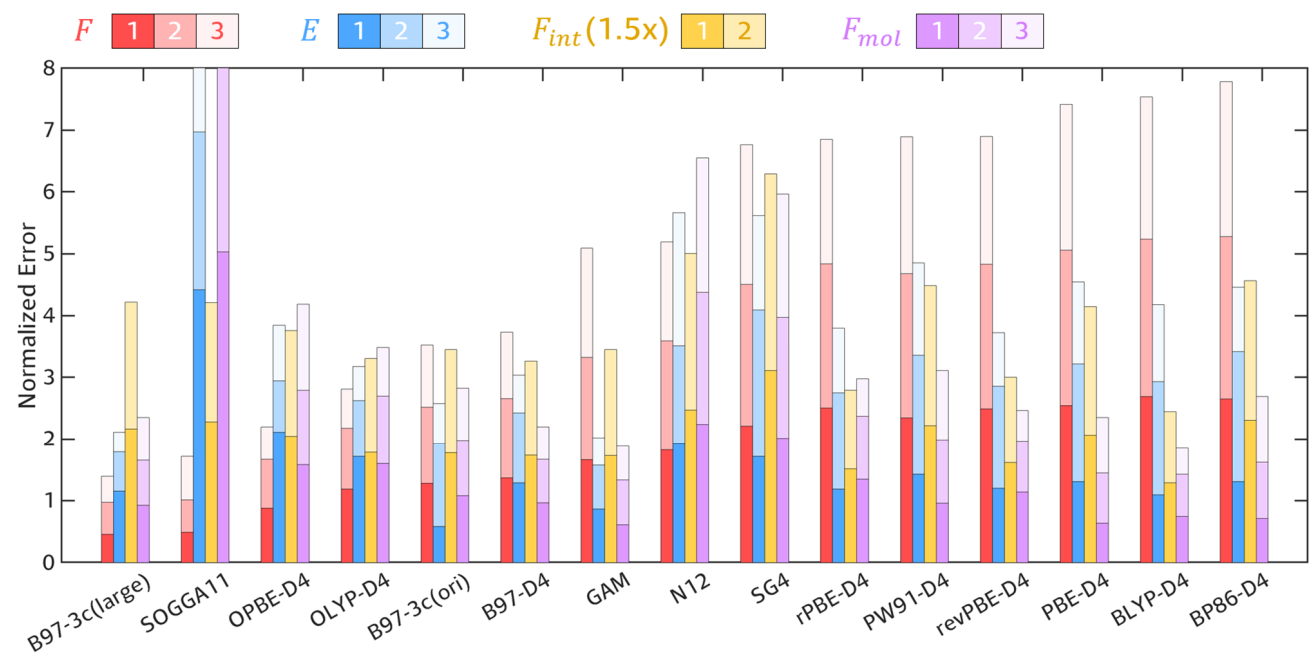

*Figure 3. The normalized error of GGA functionals in the benchmark. Each color represents a particular metric on a particular level (1/2/3) of cluster size, as shown in the legend at the top. Very high bars going beyond the range of the figure are truncated. The normalized error of $F_{int}$ is multiplied by 1.5 to compensate for the absence of layer 3, so its stacked height is comparable to other metrics.*

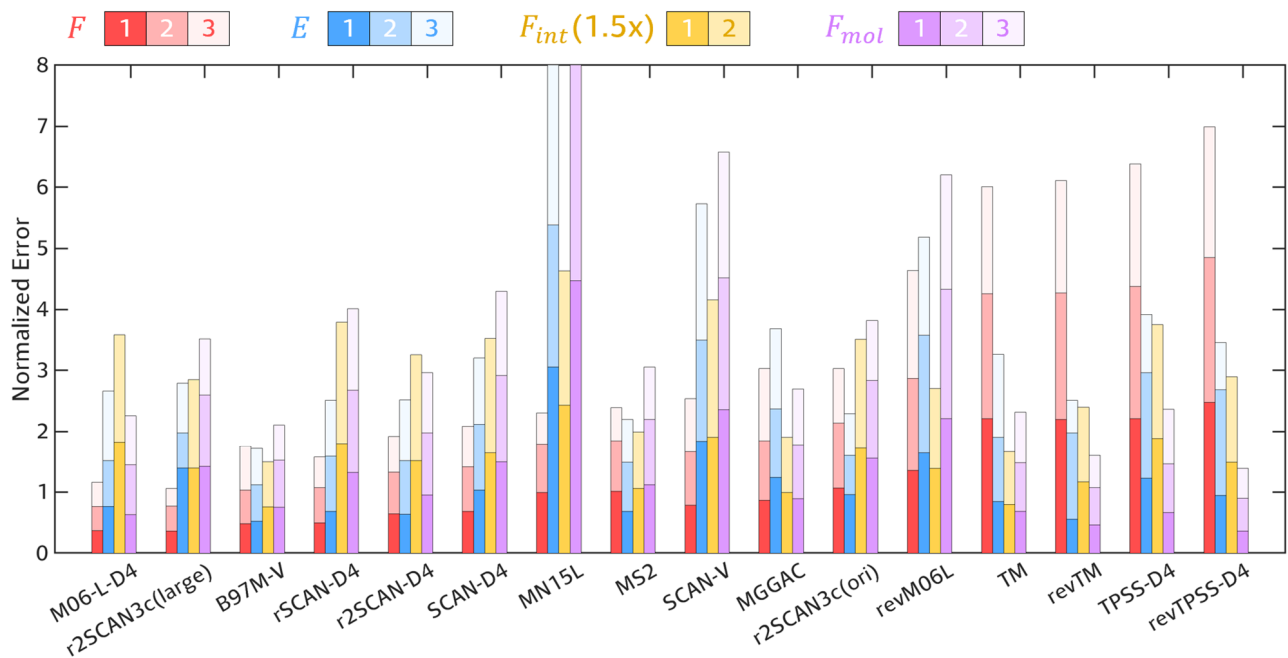

*Figure 4. The normalized error of mGGA functionals in the benchmark. See caption notes of Figure 3 for details.*

GGA functionals, due to their low computational cost, are commonly used in ab-initio MD. However, as shown in Fig. 3, no single GGA functional has very good accuracy on all metrics. The popular PBE and BLYP functionals have extremely large errors on the total force, and fairly large errors on the total energy. This could probably be attributed to the tendency of overestimating bond lengths for GGA/mGGA[116]. The B97-3c method with large basis sets shows the lowest force error in all GGAs, but its error on $F_{int}$ is large. Some other functionals, like OLYP and B97, stay in the middle. It seems that there are no overwhelmingly good options at this rung, probably due to the restricted functional form.

This situation is somewhat improved at the mGGA level (Fig. 4). Particularly, the B97M-V functional has a satisfying low error on all metrics. M06L-D4 and r2SCAN3c have even lower total force errors, but show larger errors on intermolecular forces. The MS2 functional has an impressively low error considering its simple form and low popularity. The similar may be said of MGGAC. The (rev)TM and (rev)TPSS functionals have extremely large total force errors, which may be related to their shared correlation part. The revM06L functional has surprisingly large errors, showing that revision is not always better.

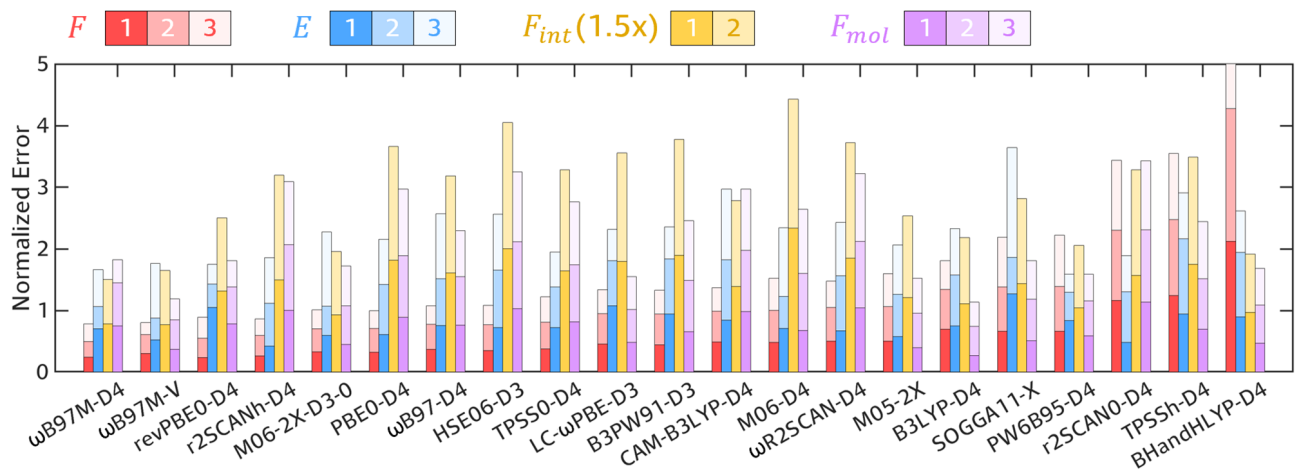

*Figure 5. The normalized error of hybrid functionals in the benchmark.*

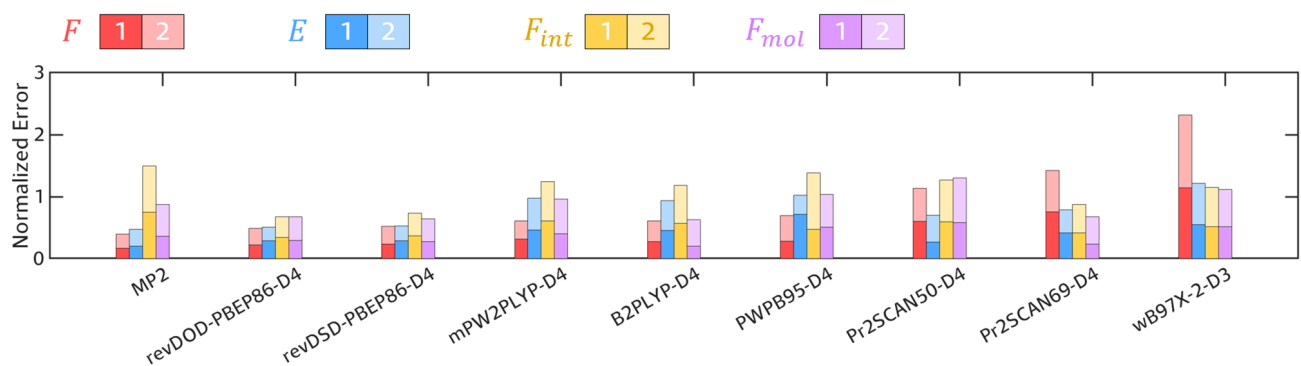

*Figure 6. The normalized error of double-hybrid functionals in the benchmark. In this figure, the normalized error of $F_{int}$ is not multiplied by 1.5, because other metrics do not have layer 3 either.*

Further general improvement is observed at rung 4 (Fig. 5). The ωB97M-V functional has the lowest error in this rung on almost all metrics. Its DFT-D counterpart ωB97M-D4 does a similar job, except for a larger error on $F_{mol}$. M06-2X outperforms all other Minnesota functionals tested, agreeing with a previous MD-based benchmark[17]. revPBE0 is significantly better than the original PBE0 on all metrics. The other classic B3LYP, is not the best one but the error is still well acceptable. The LC-ωPBE-D3 functional, which was best in an energy-based benchmark with large clusters[7], is also doing well here in the energies of large clusters, only beaten by revPBE0-D4 and PW6B95-D4 (see raw data in the supplementary files). However, as seen in Fig. 5, its overall accuracy is not exceptional. This reflects the importance of force metrics, as well as a benchmark set covering different system sizes.

The most accurate results unsurprisingly come from rung 5 (Fig. 6). On this level, only data on small and medium clusters are available, since the reference data on large clusters are also on the double-hybrid level (revDOD-PBEP86-D4). Interestingly, the lowest error of total force and energy comes from MP2 itself. Indeed, MP2 has been used once as reference energies of small clusters, due to its small error and low computational cost compared to CCSD(T)[9]. However, its error on intermolecular forces is significantly larger than the best double-hybrids. The revDSD-PBEP86-D4 and its DOD variant overwhelmingly outperform all other functionals tested, with negligible differences between the two. In case the opposite-spin only MP2 can be accelerated by the Laplace transform method[117, 118], the DOD variant should be preferred.

## 4 Discussion

### 4.1 The effect of dispersion corrections

In the results listed above, most functionals are tested with the DFT-D3/D4 correction when applicable. Even though this should be a general improvement for weak interactions, it is not rare that adding it leads to larger errors in specific cases. In this section, several functionals are re-calculated with their D4 corrections removed. The results are in Table 4.

| Metric | $F$ | | | $E$ | | | $F_{int}$ | | $F_{mol}$ | | | $E_{int}$ |
|---|---|---|---|---|---|---|---|---|---|---|---|---|
| Layer | 1 | 2 | 3 | 1 | 2 | 3 | 1 | 2 | 1 | 2 | 3 | 1 |
| r2SCAN | 0.090 | 0.098 | 0.090 | 2.02 | 2.21 | 1.39 | 0.036 | 0.060 | 0.019 | 0.028 | 0.025 | 2.88 |
| +D4 | 0.090 | 0.098 | 0.088 | 2.42 | 2.69 | 3.05 | 0.036 | 0.059 | 0.020 | 0.030 | 0.028 | 7.77 |
| revPBE0 | 0.039 | 0.050 | 0.048 | 7.94 | 3.30 | 6.34 | 0.037 | 0.048 | 0.037 | 0.040 | 0.031 | 44.73 |
| +D4 | 0.033 | 0.045 | 0.051 | 3.97 | 1.17 | 0.97 | 0.031 | 0.040 | 0.017 | 0.018 | 0.012 | 16.66 |
| mPW2PLYP | 0.044 | 0.042 | - | 1.92 | 1.43 | - | 0.023 | 0.034 | 0.007 | 0.014 | - | 2.67 |
| +D4 | 0.044 | 0.042 | - | 1.76 | 1.58 | - | 0.022 | 0.032 | 0.009 | 0.016 | - | 9.77 |

Table 4. The (raw) error of three functionals with and without D4 correction. The unit of force is eV/Å, and the unit of energy is meV. Numbers at least 10% lower than their counterparts are underlined. For the definition of $E_{int}$, see formula (9)(10) and discussions below.

The first thing to notice is that DFT-D is only a small correction to total forces. Therefore, adding or removing DFT-D would not suddenly make a functional bad at total forces good. This is understandable since a large part of the total force comes from short-range bonded interactions, while the dispersion force is only proportional to $r^{-7}$ or lower. The same can be said for $F_{int}$ as well. For $F_{mol}$, the relative change is larger due to the smaller absolute magnitude of this metric, and the complete exclusion of bonded forces. The energies, being the accumulative effect of forces, observe the largest differences.

The impact of DFT-D on accuracy is mixed. A large improvement is found for revPBE0 after D4 is added, a small deterioration is found for r2SCAN, and the numbers only change slightly for mPW2PLYP. The deterioration of r2SCAN can be explained as the original functional was already designed with dispersion effects in mind, so adding further corrections is not necessarily good. This is also reflected in the lower parameter $s_8 = 0.6018$ for r2SCAN instead of 1.5718 for revPBE0. For mPW2PLYP the correction is even smaller ($s_6 = 0.75$ and $s_8 = 0.4579$), which is expected since MP2 also partially describes the dispersion effects.

By intuition, the DFT-D correction should be small when the original functional already considers dispersion effects. This is also confirmed by the results above. Therefore, the potential "over-correction" from DFT-D should generally be smaller than the potential improvement. Along with the observation that DFT-D only has noticeable impacts on 2 of 4 metrics, the recommendation of functionals in this work should not be strongly affected by it.

In this work, only forces and relative energies are considered. Another metric that DFT-D could have a larger impact on is the total interaction energy. This quantity can be defined, for example, as the energy of the cluster minus that of individual molecules at their respective geometries:

$$E_{c,int} = E_c - \sum_{1 \leq n \leq N} E_{c,n,iso} \tag{9}$$

$$MAE(E_{int}) = \frac{\sum_{1 \leq c \leq N_c} |E_{c,int}^{ref} - E_{c,int}^{DFT}|}{N_c N} \tag{10}$$

Where $E_c$ is the total energy of the cluster $c$, $E_{c,n,iso}$ is the energy of molecule $n$ of cluster $c$ when isolated, and the summation includes all clusters at a given level. This may be considered the "dissociation energy" of the cluster, but with all geometries unrelaxed.

This quantity is difficult to calculate accurately with atomic basis sets due to the large basis set superposition error (BSSE), and it is less relevant for common applications where relative energies are of concern. Therefore, it is not primarily considered in the benchmark. As shown in Table 4, this quantity has a strong dependence on DFT-D. This can be explained as the total energy is related to the absolute depth of the DFT-D potential well, and it does not cancel at subtraction like the relative energy defined in (8). The numbers of $E_{int}$ on other functionals at level 1 are available in the supplementary files.

## 4.2 Relation between force and energy metrics

As shown in the benchmark results, low errors on forces do not always mean low errors on energies. However, since the force is the negative gradient of PES, it is expected that the two metrics are at least somewhat related. Fig. 7 shows the scatter plot between the total energy error and three force error metrics. Notably, the total energy error is somewhat correlated with the intermolecular force error $F_{int}$ and $F_{mol}$, but not the total force error.

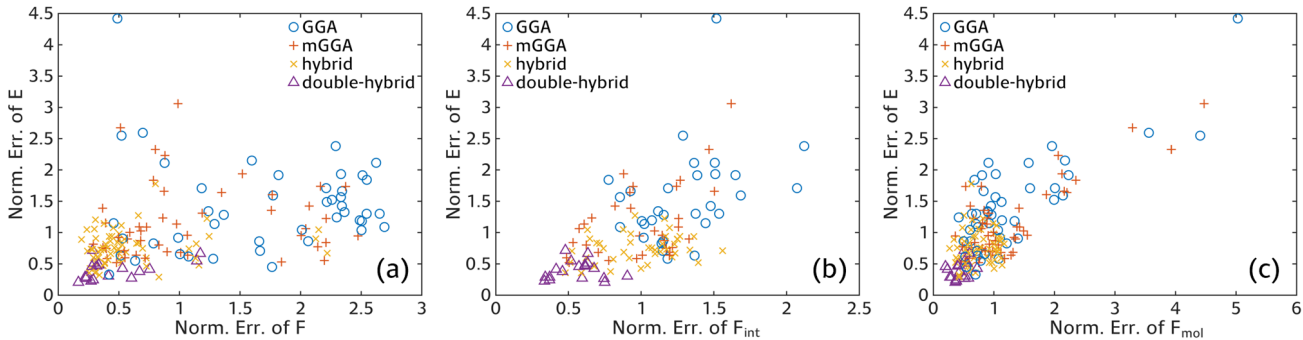

*Figure 7. The relation between the normalized error of energy and three force metrics: (a) F, (b) $F_{int}$, (c) $F_{mol}$. Each data point represents a functional at a particular layer. The colors and shapes represent the rung of the functional, as shown in the legends.*

This is understandable considering how the bonded interactions enter final force and energy metrics. In typical MD simulations, water molecules vibrate around their equilibrium positions, and the potential energy is roughly harmonic to the displacements. Since the displacements are small, their energy contributions are even smaller. This contribution may further cancel when calculating energy differences, if different clusters are vibrating at similar magnitudes. As a result, the bonded interaction only enters $MAE(E)$ to a limited extent. The total force, on the other hand, is directly proportional to the vibration displacements; and no cancellations happen when evaluating formula (3). Therefore, the total force has larger contributions from bonded interactions. This explains why $MAE(E)$ decorrelates with the total force but is more related to intermolecular forces.

It should be noted that, however, the energy and intermolecular forces are only loosely related, and they are not replacements for each other. Indeed, the correlation is barely observable if only hybrid or double-hybrid functionals are considered (the yellow and purple marks in Fig. 7). Furthermore, as shown in the next section, corrections to bonded interactions could also lead to a significant improvement in $MAE(E)$. In other words, the intramolecular contribution to energy is also not negligible. The standard of a good functional should still be low errors on both energy and forces.

## 4.3 Could GGA be improved?

As shown in Section 3, GGA functionals are generally not satisfactory in one or more metrics. A uniformly good performance requires mGGA or higher levels. However, the GGA functionals are still popular due to their simpler forms and lower computational costs. Is it possible to improve the accuracy of GGA without increasing the cost much?

As discussed above, the error of DFT can be decomposed into bonded and non-bonded parts. The non-bonded part includes dispersion and various types of electrostatic forces, which are non-trivial to model. The bonded part, however, is very simple for water with only 3 degrees of freedom. The bonded part is also what many GGA functionals struggle with, reflected by the large error of total forces. Therefore, it should be possible to improve the accuracy of GGA by simply adding corrections to bonded interactions.

In this section, this scheme is tried on the BLYP-D4 functional. The BLYP-D4 is chosen because it already does well on intermolecular forces (Fig. 3); therefore, it may become an all-rounded good method for water after the correction is applied. To find out the correction needed, the potential energy is scanned over the H-O bond length of a single water molecule, and the result is shown in Fig. 8.

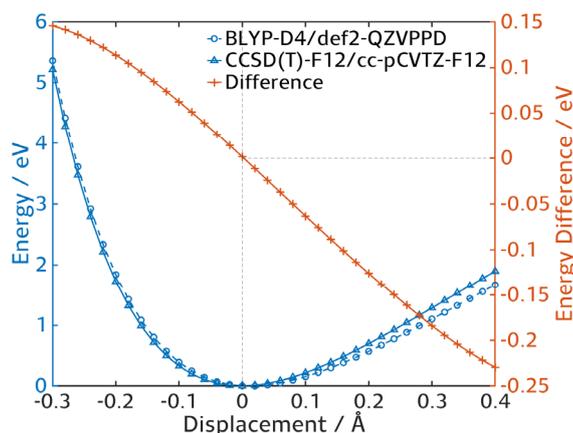

Figure 8. The single-molecule potential energy as a function of the O-H bond length. Only one O-H bond is scanned; the other bond and the angle stay at minimum positions (optimized with ωB97M-V, where O-H = 0.9587 Å, H-O-H = 104.93°). The energies and displacements are relative to the minimum point.

In simulations without bond breaking, the O-H bond typically vibrates around the minimum location. In such cases, the bond length variation can hardly be larger than -0.2~+0.3 Å, limited by the required energy of ~1 eV or ~40 kT at room temperatures. As shown in Fig. 8, the energy difference between BLYP-D4 and CCSD(T) is almost linear in this range. This enables a simple linear correction to the energy, whose slope is fitted to be 0.6158 eV/Å. This correction is then added to all bonded O-H atom pairs, identified by the distance being smaller than 1.4 Å.

The benchmark result of BLYP-D4 with this correction is given in Table 5. Notably, the total force drastically improved on all cluster sizes, becoming one of the best one within GGA. The total energy also sees a solid improvement, while the already good intermolecular forces are unchanged. This corrected method is found to be good on all metrics and all levels, comparable to the best mGGA functional B97M-V.

| Metric | $F$ | | | $E$ | | | $F_{int}$ | | $F_{mol}$ | | |
|---|---|---|---|---|---|---|---|---|---|---|---|
| Layer | 1 | 2 | 3 | 1 | 2 | 3 | 1 | 2 | 1 | 2 | 3 |
| BLYP-D4 | 0.377 | 0.366 | 0.342 | 4.12 | 5.68 | 3.77 | 0.031 | 0.040 | 0.016 | 0.020 | 0.012 |
| BLYP-D4 corrected | 0.057 | 0.057 | 0.083 | 3.09 | 1.93 | 1.88 | 0.031 | 0.040 | 0.016 | 0.020 | 0.012 |
| B97M-V | 0.067 | 0.079 | 0.107 | 1.97 | 1.84 | 1.80 | 0.018 | 0.025 | 0.016 | 0.022 | 0.017 |

Table 5. The error of BLYP-D4 with the correction. The unit of forces is eV/Å, and the unit of energies is meV. For comparison, the original BLYP-D4, as well as the overall best mGGA functional B97M-V, is also listed here.

As shown in the benchmark results, the 3c methods also improve the total force accuracy compared to the original functionals. This is likely attributed to their gCP/SRB[56] corrections, which also change the intramolecular PES. Indeed, by refitting the parameters in the modified[119] gCP[120] of r2SCAN-3c[57], the BLYP-D4 functional could be corrected in a similar way (SI Section 4).

It should be emphasized that, this correction only works for the O-H bond around the minimum position. The simulation should have no bond breaking, and the bonded/non-bonded atoms must be clearly identified. The potential energy curve of bond breaking likely depends on the specific reaction, which requires extra treatments. The non-bonded O-H interactions are governed by different physics, and should be considered separately.

## 5 Conclusion

In this work, the accuracy of DFT functionals for water is examined, with a main focus on forces. A wide range of functionals on rungs 2~5 is covered, and external correction schemes are also discussed. The recommended method on each rung is summarized in Table 7. In case the recommended method is not available, or water is studied together with other substances, the readers are encouraged to check Fig. 3~6 for other choices.

| Rung | Method |
| --- | --- |
| 2 (GGA) | BLYP-D4 with the correction in Section 4.3 |
| 3 (mGGA) | B97M-V |
| 4 (hybrid) | ωB97M-V |
| 5 (double hybrid) | revDOD-PBEP86-D4 (or revDSD-) |

Table 7. The recommended method in rung 2~5, respectively.

Generally speaking, the GGA functionals are inaccurate for water. However, their accuracy may be drastically improved with a simple correction to bonded interactions. This thinking of modeling bonded and non-bonded interactions differently is central to classical force fields, but is much less common in quantum mechanics simulations. However, as shown in this benchmark, it is not rare for a DFT functional to have very different accuracy between them. This raises the question of whether such bonded corrections can be fitted to more systems and other GGA/mGGA functionals, leading to a general approach like DFT-D.

Noncovalent Interactions. *J Chem Theory Comput* **2006,** *2* (2), 364-82.

[98] Zhao, Y.; Truhlar, D. G., The M06 suite of density functionals for main group thermochemistry, thermochemical kinetics, noncovalent interactions, excited states, and transition elements: two new functionals and systematic testing of four M06-class functionals and 12 other functionals. *Theoretical Chemistry Accounts* **2008,** *120* (1), 215-241.

[99] Adamo, C.; Barone, V., Toward reliable density functional methods without adjustable parameters: The PBE0 model. *The Journal of Chemical Physics* **1999,** *110* (13), 6158-6170.

[100] Zhao, Y.; Truhlar, D. G., Design of density functionals that are broadly accurate for thermochemistry, thermochemical kinetics, and nonbonded interactions. *J Phys Chem A* **2005,** *109* (25), 5656-67.

[101] Bursch, M.; Neugebauer, H.; Ehlert, S.; Grimme, S., Dispersion corrected r2SCAN based global hybrid functionals: r2SCANh, r2SCAN0, and r2SCAN50. *The Journal of Chemical Physics* **2022,** *156* (13).

[102] Goerigk, L.; Grimme, S., A thorough benchmark of density functional methods for general main group thermochemistry, kinetics, and noncovalent interactions. *Physical Chemistry Chemical Physics* **2011,** *13* (14), 6670-6688.

[103] Peverati, R.; Truhlar, D. G., Communication: A global hybrid generalized gradient approximation to the exchange-correlation functional that satisfies the second-order density-gradient constraint and has broad applicability in chemistry. *The Journal of Chemical Physics* **2011,** *135* (19).

[104] Grimme, S., Accurate calculation of the heats of formation for large main group compounds with spin-component scaled MP2 methods. *J Phys Chem A* **2005,** *109* (13), 3067-77.

[105] Chai, J.-D.; Head-Gordon, M., Systematic optimization of long-range corrected hybrid density functionals. *The Journal of Chemical Physics* **2008,** *128* (8).

[106] Najibi, A.; Goerigk, L., DFT-D4 counterparts of leading meta-generalized-gradient approximation and hybrid density functionals for energetics and geometries. *Journal of Computational Chemistry* **2020,** *41* (30), 2562-2572.

[107] Mardirossian, N.; Head-Gordon, M., ωB97M-V: A combinatorially optimized, range-separated hybrid, meta-GGA density functional with VV10 nonlocal correlation. *The Journal of Chemical Physics* **2016,** *144* (21).

[108] Wittmann, L.; Neugebauer, H.; Grimme, S.; Bursch, M., Dispersion-corrected r2SCAN based double-hybrid functionals. *J Chem Phys* **2023,** *159* (22).

[109] Grimme, S., Semiempirical hybrid density functional with perturbative second-order correlation. *The Journal of Chemical Physics* **2006,** *124* (3).

[110] Schwabe, T.; Grimme, S., Towards chemical accuracy for the thermodynamics of large molecules: new hybrid density functionals including non-local correlation effects. *Physical Chemistry Chemical Physics* **2006,** *8* (38), 4398-4401.

[111] Goerigk, L.; Grimme, S., Efficient and Accurate Double-Hybrid-Meta-GGA Density Functionals-Evaluation with the Extended GMTKN30 Database for General Main Group